\documentclass[prd,twocolumn,superscriptaddress,nofootinbib,floatfix]{revtex4-2}

\usepackage{graphicx,color,dcolumn,booktabs,bm}
\usepackage{amsmath,amssymb}
\usepackage{txfonts}
\usepackage{leftindex}
\usepackage{ulem}
\usepackage{feynmf}   
\usepackage{slashed}  
\usepackage{cases}
\usepackage{color}
\usepackage{multirow}
\usepackage{threeparttable}
\usepackage{epstopdf}
\usepackage{enumerate}
\usepackage[dvipsnames]{xcolor}
\definecolor{CiteBlue}{RGB}{45,52,151}
\usepackage[colorlinks=true,
            citecolor=CiteBlue,
            anchorcolor=red,
            menucolor=red,
            linkcolor=CiteBlue,
           filecolor=red,
            runcolor=red,
            urlcolor=CiteBlue,
            frenchlinks=red]{hyperref}
\usepackage{appendix}
\usepackage{orcidlink}

\newcommand{\VEV}[1]{\left\langle #1 \right\rangle}

\graphicspath{{Figures/}}

\begin{document}

\title{Revisiting neutrino-assisted self-interacting dark matter}

\author{Luqi Wang}

\author{Shu-Yuan Guo}
\email{shyuanguo@ytu.edu.cn}

\author{Xuewen Liu}
\email{xuewenliu@ytu.edu.cn}

\author{Bin Zhu}
\email{zhubin@mail.nankai.edu.cn}

\affiliation{Department of Physics, Yantai University, Yantai 264005, China}

\begin{abstract}

We assess neutrino-assisted dark matter self-scattering in two scalar-mediator realizations by combining mediator-resolved potentials, numerical scattering calculations, thermal relic production, and the adopted observational constraints. In the $t$-channel case, the adopted laboratory and cosmological bounds suppress the neutrino force contribution at points reproducing the relic abundance and yielding sizable self-interactions, which are dominated by tree-level mediator exchange. Sommerfeld enhancement and early kinetic decoupling can change the relic abundance by up to 50\% in the light mediator regime. In the $s$-channel case, thermal freeze-out fixes the coupling, leaving self-scattering well below the order of magnitude required for explanation of small-scale structure of the universe. An analytic upper envelope explains this shortfall of more than eleven orders of magnitude relative to the self-interacting dark matter benchmark. 
These results establish that, under the adopted constraints and thermal relic assumptions, the neutrino force cannot account for astrophysically relevant dark matter self-scattering in either realization.

\end{abstract}

\maketitle

\section{Introduction}

Dark matter (DM) self-interactions can redistribute energy within astrophysical halos and modify their inner structure, motivating self-interacting dark matter (SIDM) as an alternative to the collisionless description~\cite{Tulin:2017ara}. Whether a particle model can provide appreciable halo self-scattering depends on both the interaction responsible for scattering and the processes that determine the DM abundance. Neutrino-mediated interactions offer a connection between these two aspects: a coupling to neutrinos can allow DM annihilation while neutrino-pair exchange generates an effective force between DM particles~\cite{Feinberg:1968zz,Hsu:1992tg,Coy:2022cpt,Zhang:2023mcv,Ferrante:2025lbs,Ittisamai:2025oxf}, known as the ``neutrino force''. 
We characterize self-scattering by the momentum-transfer cross section per unit mass, $\sigma_T/m_\chi$, and adopt $0.1$--$10~\mathrm{cm^2\,g^{-1}}$ as a benchmark range for its potential relevance to halo dynamics~\cite{Tulin:2013teo}.

Neutrino-pair exchange has been proposed as a source of DM self-interactions within an effective contact-interaction description~\cite{Orlofsky:2021mmy}. When neutrino masses can be neglected, the resulting long-distance potential scales as $1/r^5$. Its long-distance form, however, does not determine the scattering rate without specifying the interaction at shorter distances. An explicit mediator changes the potential at distances comparable to or smaller than its Compton wavelength and correlates its strength with the underlying masses and couplings~\cite{Xu:2021daf}. Depending on the realization, mediator exchange can also produce a tree-level interaction or Sommerfeld enhancement of annihilation~\cite{Coy:2022cpt}. The effectiveness of the neutrino force must therefore be assessed within the mediator framework relevant to both scattering and relic production.

The production assumptions also affect this assessment. The proposal in Ref.~\cite{Orlofsky:2021mmy} considers asymmetric DM with a repulsive neutrino-mediated potential. Here we study attractive potentials in two scalar-mediator realizations and require thermal annihilation to reproduce the observed relic abundance. This condition fixes combinations of masses and couplings that cannot be varied independently when evaluating self-scattering. Laboratory and cosmological constraints further restrict the mediator sector and the light particles coupled to the neutrino bath~\cite{Berryman:2022hds}. Our aim is to determine the viability boundaries of neutrino-assisted self-scattering under these specified thermal and observational conditions.

We consider two realizations, following the $t$- and $s$-channel terminology of Refs.~\cite{Xu:2021daf,Coy:2022cpt}. In the $t$-channel case, the scalar mediator couples separately to DM and neutrinos. The neutrino force then corrects a tree-level Yukawa potential, and a comparison of scattering with and without the loop contribution identifies the origin of any sizable self-interaction. In the $s$-channel case, the mediator couples DM directly to a neutrino. The same interaction generates the neutrino-induced box potential and permits annihilation into neutrinos. 
Imposing the thermal relic abundance fixes the coupling for each mass pair and determines the corresponding self-scattering prediction.

We adopt the mediator-resolved potentials from the literature and calculate transfer cross sections by solving the radial Schr\"odinger equation with a partial-wave decomposition~\cite{Xu:2021daf,Tulin:2013teo}. To determine the relic abundance, we account for Sommerfeld enhancement and assess the assumption of kinetic equilibrium during freeze-out. If momentum exchange becomes inefficient while velocity-dependent annihilation remains active, the DM temperature can depart from the bath temperature and alter the final abundance~\cite{Binder:2017rgn,Binder:2021bmg}. We therefore evolve the abundance and temperature using coupled Boltzmann equations (cBE) and compare with the standard number-density Boltzmann equation (nBE)~\cite{Gondolo:1990dk}. In the $s$-channel realization, we also use the Born approximation, checked against the numerical solution, to derive an analytic upper envelope for self-scattering after imposing the relic abundance condition and the adopted bounds.

The two realizations exhibit different limitations. In the $t$-channel scan, points satisfying the adopted constraints, the requirement of relic abundance, and the self-scattering benchmark receive only a negligible correction from the neutrino force; their scattering is dominated by tree-level mediator exchange. The thermal dynamics of the light mediator framework nevertheless remain important, with Sommerfeld enhancement and early kinetic decoupling (EKD) producing relic-abundance shifts approaching $50\%$ in the illustrative benchmarks. In the $s$-channel case, self-scattering scales as $\alpha_\nu^4$, whereas annihilation scales as $\alpha_\nu^2$. Combining the relic abundance relation with the perturbative coupling limit and the cosmological lower mass bound yields an analytic envelope far below the SIDM benchmark. The numerical scan confirms $\sigma_T/m_\chi<10^{-12}~\mathrm{cm^2\,g^{-1}}$ in the investigated domain. These results quantify the restrictions on neutrino-assisted self-scattering and explain how mediator constraints and thermal production limit expectations based on the long-distance force alone.

Section~\ref{sec:review} introduces the potentials and the thermal-evolution formalism. 
Section~\ref{sec:tchan} presents the $t$-channel results and the impact of the adopted constraints. 
Section~\ref{sec:schannel} develops the analytic and numerical analysis of the $s$-channel realization, 
and Section~\ref{sec:conclusion} gives our conclusions.

\section{Neutrino-induced potentials and thermal evolution}
\label{sec:review}

\begin{figure}[!htbp]
	\centering  
    \includegraphics[width=0.45\textwidth]{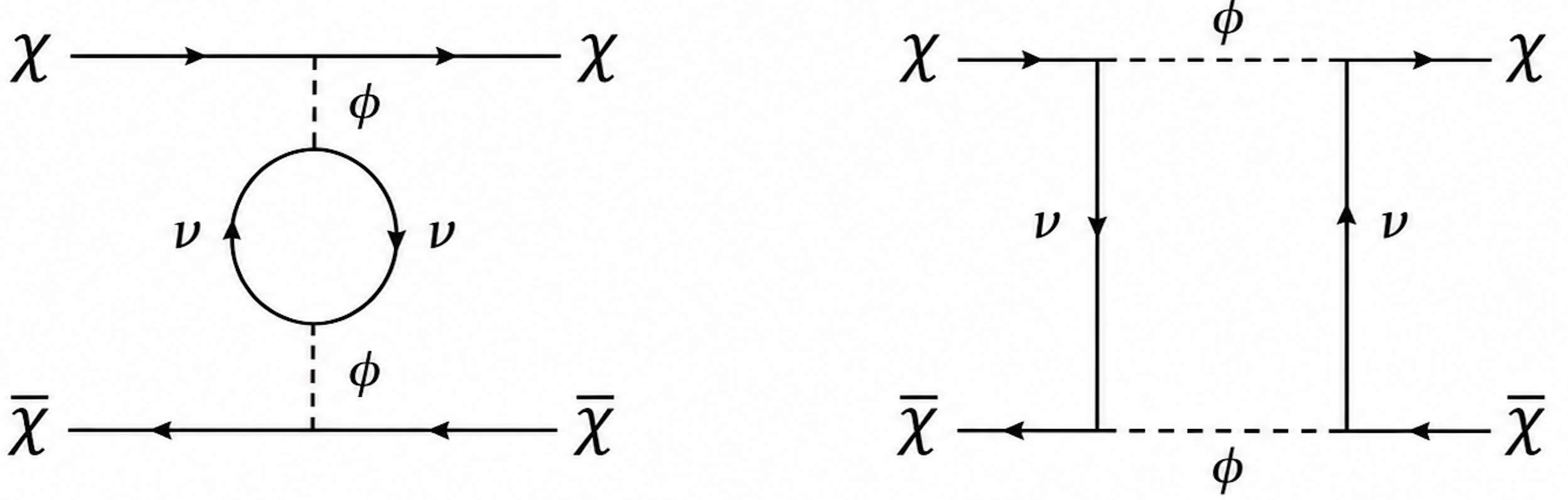}
    \caption{Feynman diagrams of $t$-channel and $s$-channel neutrino forces. \label{fig:feyn}}
\end{figure}

In the realizations considered here, neutrino-pair exchange generates a loop-induced interaction between dark matter particles, as illustrated in Fig.~\ref{fig:feyn}. We study two scalar-mediator realizations of the effective dark matter--neutrino interaction, following the $t$- and $s$-channel terminology of Ref.~\cite{Coy:2022cpt}. 

\subsection{Potentials and DM self-scattering} 
\label{subsec:potential}

In the $t$-channel realization, a scalar mediator $\phi$ couples separately to dark matter and neutrinos through the effective interaction
\begin{equation}
{\cal L}\supset - y_{\nu}\overline{\nu}\nu\phi - y_\chi\overline{\chi}\chi\phi-\frac{1}{2}m_{\phi}^{2}\phi^{2}\thinspace.\label{eq:L-t}
\end{equation}

Following the treatment of neutrino-mediated forces in Ref.~\cite{Hsu:1992tg}, the potential is obtained from the elastic-scattering amplitude in the nonrelativistic limit. 
In this realization, the neutrino force is a loop correction to the tree-level Yukawa interaction generated by $\phi$ exchange. Including both contributions gives~\cite{Xu:2021daf}
\begin{equation}
V_t(r)=\alpha_\chi\left[-\frac{e^{-m_\phi r}}{r}+\frac{\alpha_\nu}{4 \pi} m_\phi \mathcal{V}\left(m_\phi r\right)\right],
\end{equation}
where $\alpha_{\chi, \nu} \equiv y_{\chi, \nu}^2 /(4 \pi)$, and 
\begin{equation}
\mathcal{V}(x) \equiv \frac{2+(2+x) e^x \operatorname{Ei}(-x)+(2-x) e^{-x} \operatorname{Ei}(x)}{x},
\label{eq:Vt}
\end{equation}
where $\operatorname{Ei}(x) \equiv-\int_{-x}^{\infty} z^{-1} e^{-z} d z$ is the exponential integral function. 
At distances large compared with the mediator Compton wavelength, the neutrino force contribution scales as $1/r^5$.

In the $s$-channel realization, the scalar mediator couples dark matter directly to a neutrino through
\begin{equation}
{\cal L}\supset -\left(y_{\nu}\overline{\chi}\nu\phi+{\rm h.c.}\right)-\frac{1}{2}m_{\phi}^{2}\phi^{2}\thinspace,\label{eq:L-s}
\end{equation}
which generates the box diagram in Fig.~\ref{fig:feyn}. 
For simplicity, our analysis is restricted to the spin-independent component of the resulting potential~\cite{Xu:2021daf},
\begin{equation}
V_{s}\left(r\right)=-\frac{3\alpha_{\nu}^{2}}{8\pi r}\int_{0}^{\infty}\left[\frac{1}{2A}+\frac{B^{2}-tA}{4AB\sqrt{tA}}\ln\left(\frac{B-\sqrt{tA}}{B+\sqrt{tA}}\right)\right]e^{-\sqrt{t}r}dt\thinspace,\label{eq:V-s}
\end{equation}
where 
\begin{equation}
A\equiv t-4m_\chi^{2}\thinspace,\ B\equiv t+2m_{\Delta}^{2}\thinspace,\ m_{\Delta}^{2}\equiv m_{\phi}^{2}-m_\chi^{2}\thinspace.\label{eq:-16}
\end{equation}

We characterize DM self-scattering by the transfer cross section. In terms of the partial-wave phase shifts, it is given by
\begin{equation}
\sigma_T=\int d \Omega(1-\cos \theta) \frac{d \sigma}{d \Omega}=\frac{4 \pi}{k^2} \sum_{\ell=0}^{\infty}(\ell+1) \sin ^2\left(\delta_{\ell+1}-\delta_{\ell}\right) ,
\label{eq:sigmaT}
\end{equation}
where $\delta_\ell$ is the phase shift for a partial wave $\ell$. 

For the attractive, mediator-resolved potentials considered here, 
we determine the phase shifts numerically by solving the Schr\"odinger equation, 
following the partial-wave approach of Ref.~\cite{Tulin:2013teo}:

\begin{equation}
\frac{1}{r^2} \frac{d}{d r}\left(r^2 \frac{d R_{\ell}}{d r}\right)+\left(k^2-\frac{\ell(\ell+1)}{r^2}-m_{\chi} V(r)\right) R_{\ell}=0,
\label{eqn:SchrodingerEq}
\end{equation}
where $k=m_{\chi}v_{\mathrm{rel}}/2$ and $R_\ell$ is the radial wave-function. 
The phase shift $\delta_\ell$ parametrizes the asymptotic solution for $R_\ell(r)$, given by
\begin{equation} 
\lim_{r \to \infty} R_\ell(r) \propto   \cos\delta_\ell \, j_\ell(kr) - \sin\delta_\ell \, n_\ell(kr)  \, ,
\end{equation}
where $j_\ell$ ($n_\ell$) is the spherical Bessel (Neumann) function.

\subsection{Sommerfeld enhancement and EKD}
\label{subsec:thermal}

Another implication of the neutrino force is the Sommerfeld enhancement, 
which is obtained by the ratio of the wave function with and without involving long-range potential.
It is computed by solving the non-relativistic radial Schr\"odinger equation. We adopt the calculation procedure of \cite{Coy:2022cpt} in the following. For two particles of reduced mass \(\mu\) interacting through a potential $V(r)$, the relative wave function is expanded in partial waves,
\begin{equation}
   \psi(\mathbf{r})=\sum_\ell P_\ell(\cos\theta)\frac{u_\ell(r)}{r}.
\end{equation}
For potentials not more singular than $1/r$ at the origin, higher partial waves are negligible, so only $\ell=0$ is retained. 
The radial equation reads
\begin{equation}
   -\frac{1}{2\mu}u''(r)+V(r)u(r)=E u(r),
\qquad
E=\frac{k^2}{2\mu},
\end{equation}
with the physical boundary condition $u(0)=0$. 
Here $V(r)=V_t, ~V_s$ for $t$-channel and $s$-channel cases respectively. 
At large $r$, $u(r)$ must match the free solution $u_{\rm free}(r)=\sin(kr)/k$ up to a phase shift. Numerically, one solves the equation with $u(0)=0$ and $u'(0)=1$, extracts the asymptotic amplitude $A_u$, and evaluates
\begin{equation}
  S=\left|\frac{\psi(0)}{\psi_{\rm free}(0)}\right|^2
=\left|u'(0)\right|^2
=\frac{1}{k^2|A_u|^2}.  
\end{equation}

The two realizations exhibit different Sommerfeld behavior. The $t$-channel potential can produce resonant enhancement, with factors reaching $S\sim\mathcal{O}(10^4)$ in the parameter examples studied in Ref.~\cite{Coy:2022cpt}. In the $s$-channel realization, the box potential scales as $1/r^2$ at short distances and has a strength proportional to $\alpha_\nu^2$. In the parameter region considered here, it produces no appreciable Sommerfeld enhancement, consistent with Ref.~\cite{Coy:2022cpt}.


Sommerfeld enhancement can make annihilation sensitive to the dark matter velocity distribution during freeze-out. 
The standard number-density Boltzmann equation assumes kinetic equilibrium with the thermal bath. 
If elastic scattering becomes inefficient while annihilation is still active, 
the dark matter temperature can depart from the bath temperature, 
modifying the thermally averaged annihilation rate and the final abundance.

Early kinetic decoupling and its impact on relic-density calculations have been studied extensively~\cite{vandenAarssen:2012ag,Binder:2017rgn,Brummer:2019inq,Ala-Mattinen:2019mpa,Abe:2020obo,Binder:2021bmg,Zhu:2021vlz,Hryczuk:2021qtz,Abe:2021jcz,Du:2021jcj,Ala-Mattinen:2022nuj,Hryczuk:2022gay,Liu:2023kat,Aboubrahim:2023yag,Duan:2024urq,Ding:2025eqq,Abdughani:2025rsw}. 
In the light-mediator $t$-channel framework, enhanced annihilation need not be accompanied by a corresponding increase 
in the rate of momentum exchange with the bath. 
We therefore evolve the abundance and temperature together using the coupled Boltzmann equations~\cite{Binder:2017rgn,Binder:2021bmg}:
\begin{align}
\frac{Y'}{Y} &= \frac{s Y}{x \tilde H}\left[
\frac{Y_{\rm eq}^2}{Y^2} \left\langle \sigma v_{\mathrm{rel}}\right\rangle_T- \left\langle \sigma v_{\mathrm{rel}}\right\rangle_{T_\chi}
\right]\,, \label{Yfinalfinal}\\
\frac{y'}{y} &=   \frac{1}{x\tilde H}\langle C_{\text{el}} \rangle_2
+\frac{sY}{x\tilde H}\left[
\left\langle \sigma v_{\mathrm{rel}}\right\rangle_{T_\chi}-\left\langle \sigma v_{\mathrm{rel}}\right\rangle_{2,T_\chi}
\right]
+\frac{sY}{x\tilde H} \nonumber\\ 
&\quad\times\frac{Y_{\rm eq}^2}{Y^2}\left[
\frac{y_{{\rm eq}}}{y}\left\langle \sigma v_{\mathrm{rel}}\right\rangle_{2,T}\!-\!\left\langle \sigma v_{\mathrm{rel}}\right\rangle_T
\right]
+2(1-w)\frac{H}{x\tilde H}\,,\label{eq:cBE}
\end{align}
where $x\equiv m_\chi/T$ is the inverse-temperature variable, 
$Y_{\rm eq}(x)\equiv n_{\rm eq}(T)/s$, and 
\begin{equation}
\tilde H\equiv H/\left[1+ (1/3)d(\log g^s_{\rm eff})/d(\log T)\right], 
\end{equation}
with $g^s_{\rm eff}$ being the entropy degrees of freedom of the 
background plasma. 
$w(T_\chi)\equiv 1-{\langle p^4/E^3 \rangle_{T_\chi}}/({6T_\chi})$, with $\langle p^4/E^3\rangle
=\frac{g_\chi}{n_\chi^{\rm eq}(T_\chi)}\int \frac{d^3p}{(2\pi)^3} {\left( \vec{p} \cdot \vec{p} \right)^2}/{E^3} e^{- \frac{E}{T_\chi}}$. 
See Appendix~\ref{app:cBE} for more details.

\section{Effectiveness of the $t$-channel force}
\label{sec:tchan}

\subsection{Self-scattering and the relic abundance}

For the numerical calculation, we use the parameters $a\equiv v_{\mathrm{rel}}/(2\alpha_\chi)$, $b\equiv\alpha_\chi m_\chi/m_\phi$,  and introduce the reduced radial function and dimensionless radial coordinate
\begin{equation}
\chi_{\ell}\equiv rR_{\ell},\qquad x_r\equiv\alpha_\chi m_\chi r.
\end{equation}
The radial coordinate $x_r$ is distinct from the inverse-temperature variable $x=m_\chi/T$ used in the Boltzmann equations.
The radial Schr\"odinger equation, Eq.~\eqref{eqn:SchrodingerEq}, then becomes
\begin{equation}
\left(\frac{d^2}{d x_r^2}+a^2-\frac{\ell(\ell+1)}{x_r^2}   +\frac{1}{x_r} e^{-x_r / b}+\mathcal{V}_b\right) \chi_{\ell}(x_r)=0, 
\label{eq:schEq_t}
\end{equation}
where the neutrino-force contribution is
\begin{align}
    \mathcal{V}_b&=-[\alpha_{\nu}e^{-x_r/b}(2b~e^{x_r/b}+e^{2x_r/b}(2b+x_r) \operatorname{Ei}(-x_r/b)\nonumber \\
   &\quad+ (2b-x_r)\operatorname{Ei}(x_r/b))]/({4\pi b x_r}).
\end{align}

We obtain $\sigma_T$ by solving the radial equation and summing the partial-wave contributions. 
In $t$-channel, contributions from higher partial waves are not a priori negligible, 
starting from $\ell=0$, we increase $\ell_{\max}$ until the transfer cross section changes by less than $1\%$. 
We use a representative relative velocity $v_{\mathrm{rel}}=10~\mathrm{km/s}$ and 
compare $\sigma_T/m_\chi$ with the benchmark range relevant to dwarf-galaxy self-interactions 
(around $1 \mathrm{cm^2\,g^{-1}}$).

\begin{figure}[!htbp]
\includegraphics[width=0.5\textwidth]{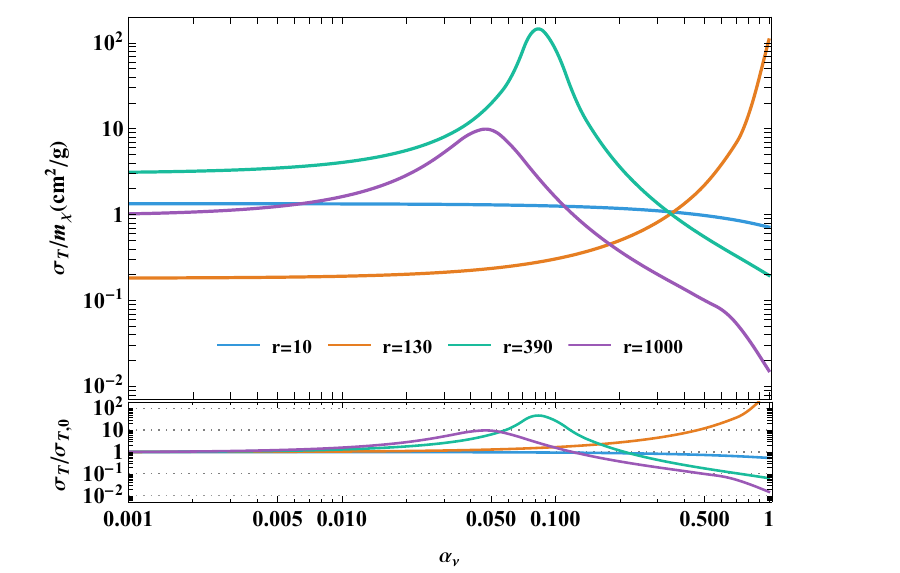}
    \caption{The $t$-channel case. \textit{Upper panel}: the
momentum-transfer self-scattering cross section per unit DM mass,
$\sigma_T/m_\chi$, as a function of the mediator--neutrino coupling
$\alpha_\nu$, with the neutrino force contribution included, for four
benchmark mass ratios $r \equiv m_\chi/m_\phi = 10$, $130$, $390$,
and $1000$, with $m_\phi$ fixed at $0.1$~GeV. 
\textit{Lower panel}: the
ratio $\sigma_T/\sigma_{T,0}$, where $\sigma_{T,0}$ is computed
without the neutrino force, quantifying the enhancement (ratio $>1$)
or suppression (ratio $<1$) of the self-interaction due to the
neutrino force contribution.}
    \label{fig:tchalphanuxsec}
\end{figure}

Figure~\ref{fig:tchalphanuxsec} shows the dependence on $\alpha_\nu$ for $m_\phi=0.1~\mathrm{GeV}$ and four mass ratios, 
$r\equiv m_\chi/m_\phi=10$, $130$, $390$, and $1000$. 
The lower panel isolates the effect of the neutrino force through the ratio $\sigma_T/\sigma_{T,0}$, 
where $\sigma_{T,0}$ is computed with the tree-level Yukawa potential alone. 
The neutrino force contribution can either enhance or suppress the total cross section. 
For instance, for $m_\chi/m_\phi =390$ (the green curve), an enhancement factor of $\sim 50$ is attainable, nearing $\alpha_\nu\sim 0.1$. 
The resonant peaks reflect the sensitivity of scattering to the bound-state structure of the combined potential, 
which changes with the coupling and mass ratio.

These benchmarks illustrate the possible impact of the neutrino force before the constraints on $\alpha_\nu$ are imposed. 
Whether a sizable effect remains compatible with both the relic abundance and the adopted bounds is assessed below.

The freeze-out calculation includes the annihilation channels $\chi\bar{\chi}\to\phi\phi$ and $\chi\bar{\chi}\to\nu\nu$. 
We also include elastic scattering, $\chi\phi\to\chi\phi$ and $\chi\nu\to\chi\nu$, 
which controls momentum exchange with the thermal bath and allows us to assess kinetic decoupling.

We calculate the relic abundance using both the 
nBE~\cite{Gondolo:1990dk} and the cBE treatment. 
We numerically solve Eqs.~\eqref{Yfinalfinal} and~\eqref{eq:cBE} for a given set of parameters ($m_\chi, m_\phi,  \alpha_\chi$). 
For simplicity, we set $\alpha_\nu = \alpha_\chi$. 
The viable parameter space is determined by matching the observed relic density to the reported $\Omega h^2 = 0.120\pm 0.001$~\cite{Planck:2018vyg}.

\begin{figure}[!htbp]
\includegraphics[width=0.425\textwidth]{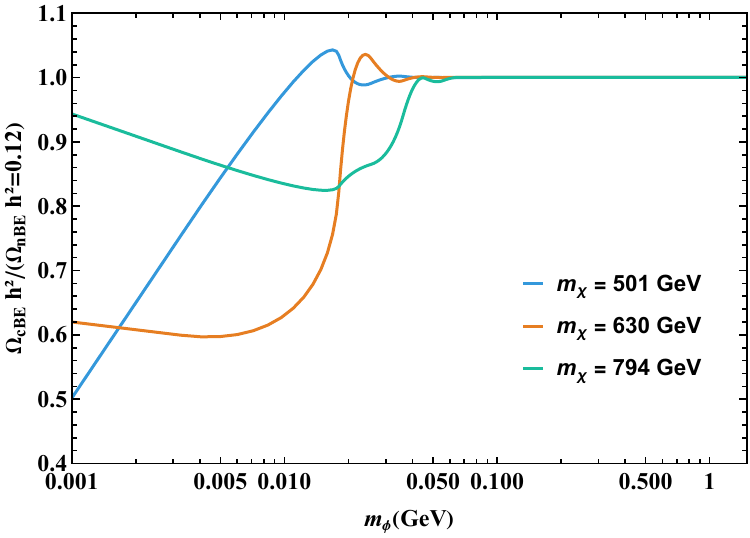}
\caption{Ratio of the relic abundances obtained from the cBE and nBE treatments, 
as a function of $m_\phi$, for $m_\chi=501$, $630$, and $794~\mathrm{GeV}$. 
The nBE abundance is fixed to $\Omega_{\rm nBE}h^2=0.12$, 
and the cBE abundance is evaluated at the same parameter point. 
The cBE treatment follows the dark matter temperature as well as its abundance. 
Deviations from kinetic equilibrium earlier modify the relic abundance by up to 
approximately $50\%$ in the light-mediator benchmarks shown, 
whereas the two calculations agree for $m_\phi\gtrsim0.1~\mathrm{GeV}$.}
    \label{fig:tnbecBE}
\end{figure}

Figure~\ref{fig:tnbecBE} compares the two treatments for $m_\chi=501$, $630$, and $794~\mathrm{GeV}$ as a function of $m_\phi$, 
to highlight the significance of the improved treatment of decoupling dynamics near the enhanced region. 
At each point, the parameters are chosen to give $\Omega_{\rm nBE}h^2=0.12$, and the cBE result is evaluated for the same parameters. 

For $m_\phi\gtrsim0.1~\mathrm{GeV}$ in these benchmarks, 
the two calculations agree, since the interaction reduces to a contact vertex and kinetic decoupling proceeds in the standard way. 
At lower mediator masses, however, the temperature evolution becomes important for 
Sommerfeld-enhanced annihilation, and the relic abundances can differ appreciably.
In this case, the long-range potential experienced by the
non-relativistic DM pairs Sommerfeld enhances the annihilation rate, 
the momentum transfer rate $\gamma(T)$ (see Eq.~\eqref{eq:gamma}) is strongly suppressed compared to the annihilation rate. 
This suppression arises because the small coupling $\alpha_\chi$ required to satisfy the relic density 
condition does not benefit from a corresponding enhancement of $\gamma(T)$, 
leading to the early decoupling of kinetic equilibrium.

The cBE result deviates from
the nBE one by as much as $\sim 50\%$ in the light mediator regime,
$m_\phi \sim \mathcal{O}(\mathrm{MeV})$. 
These deviations
demonstrate that the standard nBE treatment is inadequate for
long-range DM interactions: the parameter space reproducing the
observed relic abundance is shifted accordingly, which we take fully
into account in the scan presented below.

\begin{figure}[!htbp]
\begin{center}
    \includegraphics[width=7.5cm]{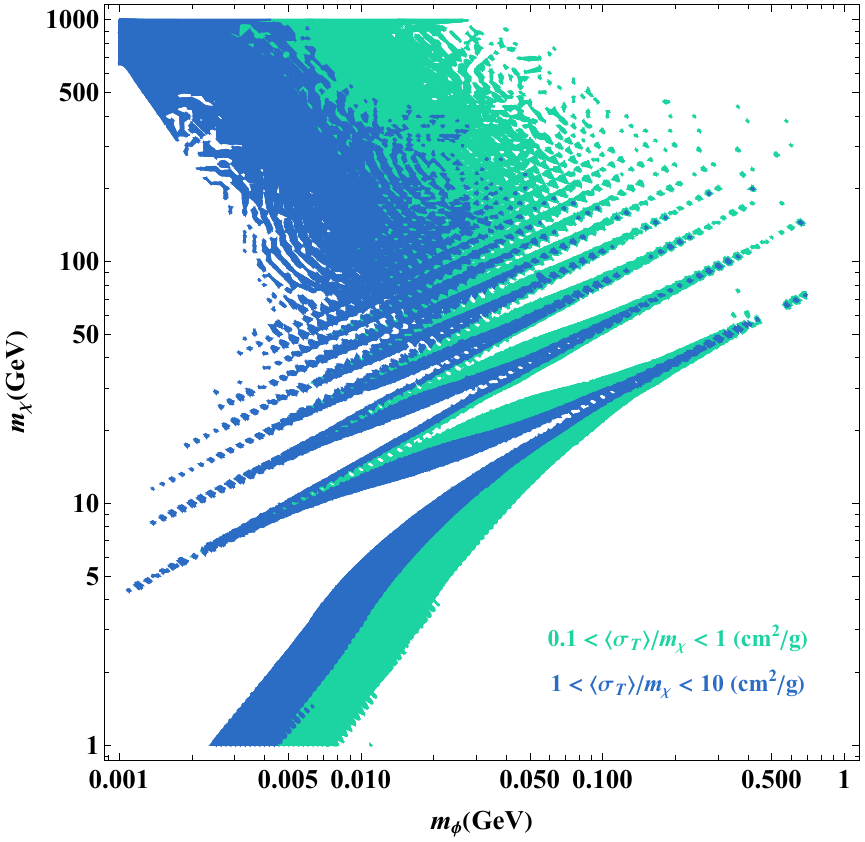}
    \caption{Viable parameter space in the $(m_\phi, m_\chi)$ plane. All displayed points reproduce 
    the observed DM relic abundance $\Omega h^2 \simeq 0.12$ within the cBE formalism. 
    The self-scattering cross section is averaged over the relative-velocity distribution specified in the text. 
    Green points yield $0.1 < \langle\sigma_T\rangle/m_\chi < 1~\mathrm{cm^2\,g^{-1}}$, 
    favored at cluster scales, while blue points yield $1 < \langle\sigma_T\rangle/m_\chi < 10~\mathrm{cm^2\,g^{-1}}$, 
    relevant for galactic and dwarf scales. }
    \label{fig:tSelf&Relic}
    \end{center}
\end{figure}

Figure~\ref{fig:tSelf&Relic} shows the parameter space that reproduce the observed relic abundance in the cBE treatment 
and yield $0.1<\langle\sigma_T\rangle/m_\chi<10~\mathrm{cm^2\,g^{-1}}$. 
Here, to account for the distribution of relative velocities in a halo, 
we average the transfer cross section over a Gaussian relative-velocity distribution, 
\begin{equation}
\langle \sigma_T \rangle = \int \frac{d^3 v_{\mathrm{rel}}}{(2\pi v_0^2)^{3/2}} e^{- v_{\mathrm{rel}}^2/(2v_0^2)} \sigma_T(v_{\mathrm{rel}}),
\end{equation}
where $v_{\mathrm{rel}}$ denotes the relative speed and $v_0$ represents the most probable velocity, 
set here to $v_0 = 10~\mathrm{km/s}$. 
The green region,
$0.1$--$1~\mathrm{cm^2\,g^{-1}}$, is favored by cluster-scale observations, whereas the blue region,
$1$--$10~\mathrm{cm^2\,g^{-1}}$, addresses the small-scale
anomalies of galactic and dwarf halos. 

The banded structure reflects the resonant dependence of the long-range dynamics on the model parameters, 
which affects both self-scattering and the coupling required by the relic abundance. 
The displayed points extend from dark matter masses of a few GeV to the TeV scale. 
They establish the overlap between the relic-abundance condition and the adopted self-scattering benchmark. 
The additional constraints on the mediator sector and the specific contribution of the neutrino force are examined in the next subsection.

\subsection{Neutrino force contribution with constraints}

\begin{figure}[!htbp]
\begin{center}
    \includegraphics[width=7.85cm]{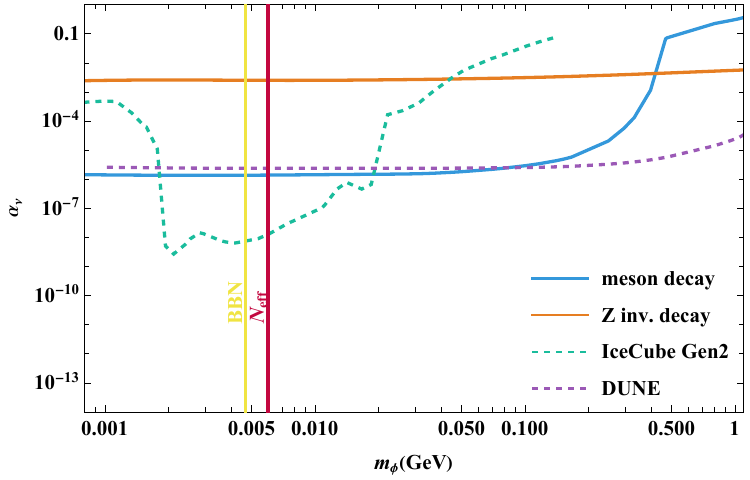}
    \caption{Experimental and observational constraints on the mediator--neutrino coupling $\alpha_\nu$ as 
    a function of the mediator mass $m_\phi$. The solid blue and orange lines are the upper bounds 
    from rare meson decays and from the invisible decay width of the $Z$ boson, respectively. 
    The meson decay bound is essentially flat, $\alpha_\nu \lesssim 2\times10^{-6}$, 
    for $m_\phi$ below the pion and kaon thresholds, 
    and degrades rapidly once the decay channels close kinematically, 
    while the $Z$-width bound is mass-independent at $\alpha_\nu \simeq 10^{-3}$ for $m_\phi \ll m_Z$. 
    The dashed green and purple lines show the projected sensitivities of IceCube-Gen2 and DUNE, respectively. 
    The vertical yellow and dark-red lines mark the lower bounds on $m_\phi$ from BBN 
    and from the effective number of relativistic degrees of freedom $N_{\rm eff}$, respectively.}
    \label{fig:t-constraint}
    \end{center}
\end{figure}

We now examine whether the neutrino force contribution can remain significant after imposing constraints on the mediator sector. 
The mediator--neutrino coupling induces neutrino self-interactions and is subject to laboratory, astrophysical, and cosmological bounds~\cite{Berryman:2022hds}.
A light mediator can enhance dark matter self-scattering, but its thermal population also affects the neutrino bath. 
Under the thermal assumptions adopted here, entropy transfer from $\phi$ to neutrinos increases $N_{\rm eff}$. 
The adopted $N_{\rm eff}$ constraint excludes mediator masses below approximately $6~\mathrm{MeV}$~\cite{Planck:2018vyg}; 
the corresponding lower mass limit is shown in Fig.~\ref{fig:t-constraint}.

The additional energy density carried by a light mediator modifies the expansion rate during Big Bang nucleosynthesis (BBN). 
This changes the freeze-out of neutron--proton interconversion and the subsequent evolution of the primordial light-element abundances. 
We evaluate the helium and deuterium abundances using \texttt{PRyMordial}~\cite{Burns:2023sgx} 
and include the resulting BBN mass bound in Fig.~\ref{fig:t-constraint}.

Laboratory probes include rare meson decays and the invisible decay width of the $Z$ boson. 
The meson-decay bounds depend on the neutrino flavor structure and are particularly restrictive 
for couplings involving $\nu_e$ or $\nu_\mu$. High-energy neutrino observations provide complementary 
probes through scattering on the cosmic neutrino background. 
Figure~\ref{fig:t-constraint} shows the adopted laboratory bounds as solid curves and the projected sensitivities 
of IceCube-Gen2 and DUNE as dashed curves. These projections indicate possible improvements 
in sensitivity and are distinguished from the existing bounds used in the analysis.

The solid blue curve gives the adopted meson-decay bound. For mediator masses well below the relevant decay thresholds, 
it is approximately $\alpha_\nu\lesssim2\times10^{-6}$ and weakens as the available phase space closes. 
The solid orange curve gives the bound from the invisible $Z$ width, approximately $\alpha_\nu\lesssim10^{-3}$ for $m_\phi\ll m_Z$. 
In the range shown, the latter becomes the dominant laboratory bound at mediator masses of order $0.4~\mathrm{GeV}$ and above.

\begin{figure}[!htbp]
\begin{center}
    \includegraphics[width=7.5cm]{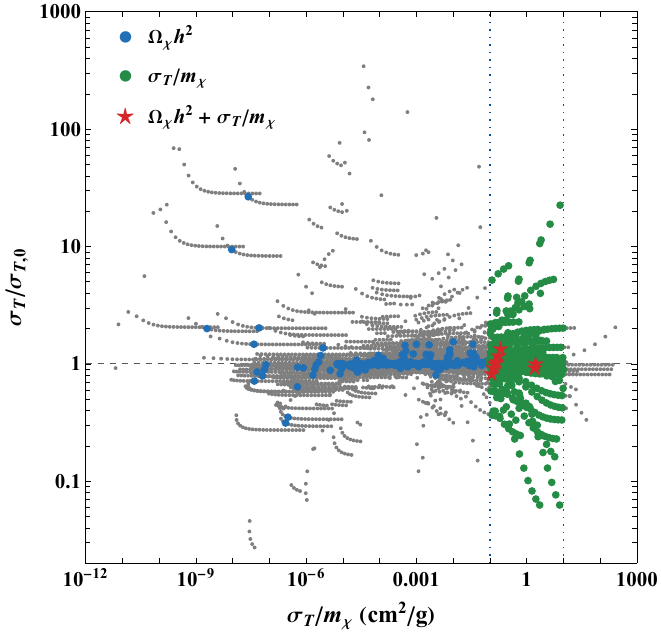}
    \caption{Parameter points satisfying the adopted bounds on $\alpha_\nu$ in Fig.~\ref{fig:t-constraint}. 
    The horizontal axis is the total momentum-transfer cross section per unit mass, $\sigma_T/m_\chi$, 
    including the neutrino-force correction; the vertical axis is $\sigma_T/\sigma_{T,0}$, 
    where $\sigma_{T,0}$ is obtained from the tree-level Yukawa potential alone. 
    Gray points show the retained scan, blue points satisfy the relic-abundance requirement, 
    and green points lie within $0.1$--$10~\mathrm{cm^2\,g^{-1}}$, marked by the vertical dotted lines. 
    Red stars satisfy both requirements. Their proximity to the horizontal line $\sigma_T/\sigma_{T,0}=1$ 
    shows that self-scattering at these points is dominated by tree-level mediator exchange.}
    \label{fig:t-effect}
    \end{center}
\end{figure}

We next quantify the impact of the adopted bounds on neutrino-assisted self-scattering. 
To this end, we perform a scan over the model parameter space, retaining only those points that respect the above experimental constraints, 
and compute for each point the self-scattering cross section both with ($\sigma_T$) and without ($\sigma_{T,0}$) the neutrino-force contribution. 
The ratio $\sigma_T/\sigma_{T,0}$ measures the change caused by the neutrino force contribution. 
Figure~\ref{fig:t-effect} displays this ratio against the total self-scattering strength $\sigma_T/m_\chi$, which makes the consequence of
the $\alpha_\nu$ bounds immediately transparent.

The decisive feature of Fig.~\ref{fig:t-effect} is that the red stars, 
which satisfy both the relic-abundance requirement and the self-scattering benchmark, 
lie close to $\sigma_T/\sigma_{T,0}=1$. Other points can exhibit larger relative changes, 
but they do not simultaneously satisfy these two conditions. 
Thus, under the adopted constraints, the sizable self-scattering found in the 
viable $t$-channel points is dominated by tree-level mediator exchange, 
with only a negligible correction from the neutrino force. 
This demonstrates that, contrary to naive expectations, 
sizable neutrino-assisted self-interactions are not a generic outcome 
of DM--neutrino coupled models in the $t$-channel realization.


\section{The $s$-channel realization}
\label{sec:schannel}

In the $s$-channel realization, the scalar mediator couples DM directly to a neutrino through the interaction in Eq.~\eqref{eq:L-s}. This interaction generates the neutrino force through the box diagram in Fig.~\ref{fig:feyn} and permits annihilation through $\chi\bar\chi\to\nu\bar\nu$. For each mass pair ($m_\chi$,$m_\phi$), imposing the thermal relic abundance fixes the coupling and determines the corresponding self-scattering cross section.

For $m_\phi>m_\chi$, the spin-independent box potential in Eq.~\eqref{eq:V-s} is attractive and scales as $V_s\propto\alpha_\nu^2$. In the investigated region, the numerical results support the Born approximation and $s$-wave dominance, as quantified below. The leading contribution to annihilation is also $s$-wave and velocity independent,
\begin{equation}
\sigma v_{\mathrm{rel}}(\chi\bar\chi\to\nu\bar\nu)
=\frac{\pi\alpha_{\nu}^{2}m_{\chi}^{2}}{(m_{\chi}^{2}+m_{\phi}^{2})^{2}}.
\label{eq:sigmav-s}
\end{equation}
We find no appreciable Sommerfeld enhancement in the region considered, consistent with Ref.~\cite{Coy:2022cpt}. The numerical thermal-evolution calculation shows that kinetic equilibrium is maintained through freeze-out. The cBE and nBE treatments give relic abundances that agree within numerical accuracy.

To obtain an analytic description of self-scattering, we use the first Born approximation. With $k=m_\chi v_{\mathrm{rel}}/2$, the $s$-wave phase shift is
\begin{equation}
\delta_0^{\mathrm B}=-2m_\chi k\int_0^\infty
V_s(r)j_0^2(kr)\,r^2\,dr.
\label{eq:born-delta-s}
\end{equation}
In the low-velocity limit, this becomes
\begin{equation}
\delta_0^{\mathrm B}\simeq
v_{\mathrm{rel}}
\mathcal I(\delta_m),
\label{eq:born-lowv-s}
\end{equation}
where $\delta_m\equiv m_\Delta^2/m_\chi^2$. The dimensionless function $\mathcal I$ is defined by the low-velocity limit of Eq.~\eqref{eq:born-delta-s},
\begin{equation}
\mathcal I(\delta_m)\equiv
-m_\chi^2
\int_0^\infty V_s(r)\,r^2\,dr.
\label{eq:integral-function-s}
\end{equation}
This function decreases with $\delta_m$ and has the asymptotic form
\begin{equation}
\mathcal I(\delta_m)\simeq \frac{3 \alpha_\nu^2}{16 \pi} \frac{\pi^2-4}{32\delta_m},\qquad\delta_m\gg1.
\label{eq:integral-asymptotic-s}
\end{equation}
Since $\delta_0^{\mathrm B}$ is linear in $v_{\mathrm{rel}}$ at low velocity, the factor $(\delta_0^{\mathrm B})^2$ cancels the velocity dependence of $1/k^2$ in Eq.~\eqref{eq:sigmaT}. The leading $s$-wave cross section is therefore velocity independent, so halo velocity averaging leaves this result unchanged at this order. Taking the additional limit $\delta_m\gg1$ gives
\begin{equation}
\frac{\sigma_T}{m_\chi}\simeq 1.316\times10^{-6}\,\alpha_\nu^4
\left(\frac{m_\chi}{\mathrm{GeV}}\right)
\left(\frac{m_\Delta}{\mathrm{GeV}}\right)^{-4}
\mathrm{cm^2/g}.
\label{eq:master-s}
\end{equation}
The self-scattering cross section scales as $\alpha_\nu^4$, compared with the $\alpha_\nu^2$ dependence of annihilation in Eq.~\eqref{eq:sigmav-s}. We check the Born treatment by solving the radial Schr\"odinger equation~\eqref{eqn:SchrodingerEq} numerically with the same method used for the $t$-channel case. We evaluate the potential in Eq.~\eqref{eq:V-s} by numerical integration and obtain the transfer cross section from Eq.~\eqref{eq:sigmaT}, initially retaining $\ell=0$. The numerical phase shifts agree with Eq.~\eqref{eq:born-delta-s} to better than $3\%$ throughout the investigated parameter region and to better than $0.1\%$ after imposing the relic-abundance condition below. The contribution from higher partial waves is assessed separately.

We next impose the condition of DM relic abundance. Reproducing $\Omega_\chi h^2=0.12$ requires $\langle\sigma v_{\mathrm{rel}}\rangle_{\mathrm{req}} \simeq 10^{-25}~\mathrm{cm^3\,s^{-1}}$ in the thermal scenario considered. Expressing this quantity in natural units and using Eq.~\eqref{eq:sigmav-s} gives
\begin{equation}
\alpha_{\nu}^{\mathrm{relic}}(m_{\chi},m_{\phi})
=\frac{m_{\chi}^{2}+m_{\phi}^{2}}{m_{\chi}}
\sqrt{\frac{\langle\sigma v_{\mathrm{rel}}\rangle_{\mathrm{req}}}{\pi}}.
\label{eq:relicline-s}
\end{equation}
Substituting Eq.~\eqref{eq:relicline-s} into Eq.~\eqref{eq:master-s} gives the cross section per unit mass in the limit $\delta_m\gg1$ \footnote{The $\delta_m\gg1$ limit is not an extra assumption but the regime that
controls the bound: since $\sigma_T/m_\chi\propto\alpha_\nu^4$ and the relic
line gives $\alpha_\nu^{\rm relic}\propto(m_\chi^2+m_\phi^2)/m_\chi$, the
largest cross section requires the largest allowed coupling, i.e.\ the largest
$m_\phi$, this is confirmed by the numerical scan below.},
\begin{equation}
\frac{\sigma_T}{m_\chi}\simeq1.2188\times10^{-23}
\left(\frac{m_{\phi}}{\mathrm{GeV}}\right)^{4}
\left(\frac{m_{\chi}}{\mathrm{GeV}}\right)^{-3}\mathrm{cm^2\,g^{-1}}.
\label{eq:relicpred-s}
\end{equation}
Within this limit, increasing $m_\phi$ or decreasing $m_\chi$ increases $\sigma_T/m_\chi$. These changes are restricted by the adopted coupling and mass bounds. The perturbative coupling limit, $\alpha_\nu\le1$, places an upper bound on the mediator mass through Eq.~\eqref{eq:relicline-s},
\begin{equation}
\left(\frac{m_\phi}{\mathrm{GeV}}\right)^2\lesssim 1.81\times10^4\left(\frac{m_\chi}{\mathrm{GeV}}\right).
\label{eq:phicap-s}
\end{equation}
The adopted $N_{\mathrm{eff}}$ constraint imposes the lower mass bound $m_\chi>6.5~\mathrm{MeV}$ \footnote{The light state constrained here is the fermion \(\chi\), not the scalar \(\phi\) of the \(t\)-channel case. Since they differ in internal degrees of freedom and spin statistics, we have recomputed this bound with \texttt{PRyMordial}~\cite{Burns:2023sgx} for the \(s\)-channel field content, obtaining \(m_\chi \gtrsim 6.5\)~MeV.}. Combining Eqs.~\eqref{eq:phicap-s} and~\eqref{eq:relicpred-s} yields the analytic upper envelope
\begin{equation}
\frac{\sigma_{T}}{m_{\chi}}\lesssim
4.0\times10^{-15}\left(\frac{m_{\chi}}{\mathrm{GeV}}\right)^{-1}
\mathrm{cm^2\,g^{-1}},
\label{eq:bound-s}
\end{equation}
whose largest value is approached near the boundary $m_\chi=6.5~\mathrm{MeV}$, $m_\phi=10.9~\mathrm{GeV}$, and $\alpha_\nu=1$. There, $\sigma_T/m_\chi\simeq6.2\times10^{-13}~\mathrm{cm^2\,g^{-1}}$, a factor of approximately $1.6\times10^{11}$ below the lower edge of the adopted SIDM benchmark range.

\begin{figure}[!htbp]
\begin{center}
\includegraphics[width=8.6cm]{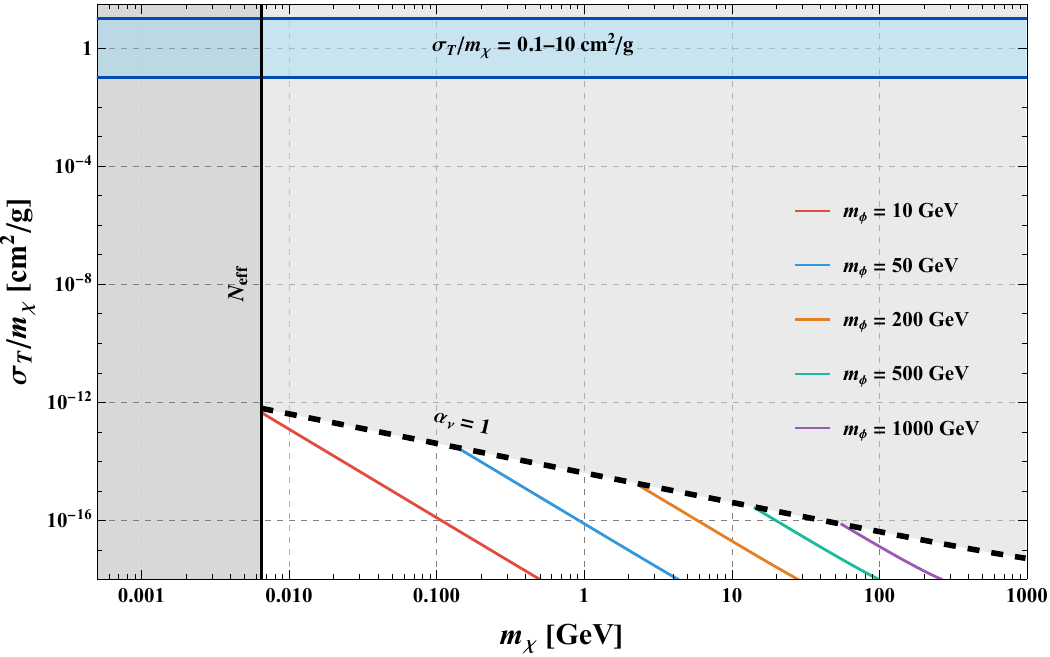}
\caption{Self-scattering cross section per unit DM mass in the $s$-channel realization after imposing the thermal relic abundance. The solid colored curves show the numerical results for five fixed mediator masses, with $\alpha_\nu=\alpha_\nu^{\mathrm{relic}}(m_\chi,m_\phi)$ from Eq.~\eqref{eq:relicline-s}. The cross section is obtained from the numerical variable-phase solution of the radial Schr\"odinger equation. The curves are displayed for $\alpha_\nu\le1$. The black dashed line shows the upper envelope at $\alpha_\nu=1$, corresponding to the largest mediator mass compatible with the relic-abundance condition at each DM mass within the adopted coupling limit. The vertical line marks the $N_{\mathrm{eff}}$ mass bound at $m_\chi=6.5~\mathrm{MeV}$, and the shaded band denotes the adopted SIDM benchmark range, $0.1$--$10~\mathrm{cm^2\,g^{-1}}$.}
\label{fig:s-excluded}
\end{center}
\end{figure}

We evaluate the numerical self-scattering prediction on a grid in $m_\chi$ and $m_\phi$, 
fixing the coupling through Eq.~\eqref{eq:relicline-s}. The scan imposes $\alpha_\nu\le1$ and $m_\chi>6.5~\mathrm{MeV}$, as in Fig.~\ref{fig:s-excluded}. 
Including the $\ell=1$ and $\ell=2$ partial waves changes $\sigma_T$ by less than $1\%$, supporting the $s$-wave approximation. 
The resulting cross section remains below $10^{-12}~\mathrm{cm^2\,g^{-1}}$ throughout the scanned region.
Figure~\ref{fig:s-excluded} shows the self-scattering cross section after imposing the relic-abundance condition for five fixed mediator masses. 
The solid curves and the dashed upper envelope at $\alpha_\nu=1$ lie well below the shaded SIDM benchmark range. 
Under these thermal and parameter assumptions, the $s$-channel neutrino force is therefore too weak to reach the adopted SIDM benchmark.

\section{Conclusion}
\label{sec:conclusion}

We have assessed neutrino-assisted DM self-scattering in two explicit scalar-mediator realizations by combining the interaction potentials, thermal relic production, and the adopted laboratory and cosmological constraints. This joint treatment identifies the conditions that limit the neutrino-force contribution and explains why a long-range neutrino-induced potential alone does not establish an astrophysically relevant self-interaction.

In the $t$-channel realization, the neutrino-loop contribution can substantially modify self-scattering when the mediator--neutrino coupling is unconstrained. After imposing the adopted bounds, the points that reproduce the observed relic abundance and reach the self-scattering benchmark instead satisfy $\sigma_T/\sigma_{T,0}\simeq1$, where $\sigma_{T,0}$ is obtained from the tree-level Yukawa potential alone. Thus, sizable self-interactions remain possible in this realization, but their dominant origin is tree-level mediator exchange. The constraints determine the physical source of the surviving scattering signal as well as the allowed parameter region.

The thermal history of the light-mediator $t$-channel framework remains important. Sommerfeld enhanced annihilation and inefficient momentum exchange can invalidate the kinetic equilibrium approximation during freeze-out. Evolving the abundance and temperature together changes the predicted relic abundance by up to approximately $50\%$ relative to the nBE treatment in the illustrative benchmarks. This result concerns the thermal dynamics of the full mediator framework; it does not require a sizable neutrino-loop contribution to the allowed self-scattering signal.

In the $s$-channel realization, fixing the thermal relic abundance removes the freedom to increase the coupling independently to enhance self-scattering. The numerical calculation supports $s$-wave dominance and the Born description in the investigated region. In the heavy-mediator limit, the relic-abundance condition gives $\sigma_T/m_\chi\propto m_\phi^4/m_\chi^3$. The perturbative coupling limit bounds $m_\phi$ from above, while the adopted $N_{\mathrm{eff}}$ constraint bounds $m_\chi$ from below. Their combination gives the analytic envelope in Eq.~\eqref{eq:bound-s}, whose largest value is approximately $6.2\times10^{-13}~\mathrm{cm^2\,g^{-1}}$ near $m_\chi=6.5~\mathrm{MeV}$, $m_\phi=10.9~\mathrm{GeV}$, and $\alpha_\nu=1$. This lies more than eleven orders of magnitude below the lower edge of the adopted SIDM benchmark. The numerical scan confirms cross sections below $10^{-12}~\mathrm{cm^2\,g^{-1}}$ throughout the studied domain. No appreciable Sommerfeld enhancement is found, and the cBE and nBE relic abundance calculations agree within numerical accuracy.

The $s$-channel conclusion applies to the spin-independent potential, thermal production scenario, and parameter domain considered, including $\alpha_\nu\le1$ and the adopted lower bound on $m_\chi$. Within these assumptions, the analytic envelope explains the numerical shortfall and excludes this realization as a source of self-scattering in the adopted benchmark range. Together, the two analyses delimit the effectiveness of neutrino-assisted SIDM in the frameworks studied: mediator constraints suppress the neutrino correction to viable $t$-channel scattering, while relic production and the mass bounds keep $s$-channel self-scattering extremely small. 

\section*{Acknowledgments}

We are grateful to  Bingrong Yu for the valuable discussions. 
The work by X.L. is supported by the Natural Science Foundation of Shandong province under the Grant No. ZR2025MS02 
and the Project of Shandong Province Higher Educational Science and Technology Program under Grants No. 2022KJ271.  
S.-Y. G. is supported by the NNSFC under Grant No. 12305113.
B.Z. is supported by NNSFC under Grant Nos. 12275232 and the Natural Science Foundation of Shandong province under the Grant No. ZR2025QA20.

\appendix

\section{Coupled Boltzmann equations}
\label{app:cBE}

To investigate the DM relic abundance with the impact of early kinetic decoupling taken into account, 
we focus on the following Boltzmann equation for the DM phase-space distribution~\cite{Binder:2017rgn,Binder:2021bmg}:
\begin{align}
 E \left( \frac{\partial}{\partial t} - H \vec{p}\cdot \frac{\partial}{\partial \vec{p}} \right) f_\chi (t, \vec{p})=
C_{\rm ann.}[f_\chi] + C_{\rm el.}[f_\chi],
\label{Eq:boltzmann}
\end{align}
where $E$ is the DM energy, $H$ is the Hubble rate, $\vec{p}$ is the DM momentum,
and $f_\chi$ is the DM phase-space density.
The collision terms, $C_{\rm ann.}$ and $C_{\rm el.}$, account for DM annihilation into thermal bath particles 
and elastic scattering between DM and SM particles, respectively.
For two-body processes, 
\begin{align}
 C_{\rm ann.}&=\frac{1}{2 g_\chi}
\sum
\int \frac{d^3 p'}{(2\pi)^3 2 E_{p'}}
\int \frac{d^3 k}{(2\pi)^3 2 E_k}
\int \frac{d^3 k'}{(2\pi)^3 2 E_{k'}}\nonumber\\
&\quad\times  
(2\pi)^4 \delta^4(p + p' - k - k') \nonumber\\
&\quad\times 
\Bigl(-|{\cal M}_{\chi \chi \to {\cal B} {\cal B}'}|^2 f_\chi(\vec{p}) f_\chi(\vec{p'}) (1 \pm f^{\rm eq}_{\cal B}(\vec{k}))\nonumber\\
&\quad\times(1 \pm f^{\rm eq}_{\cal B'}(\vec{k'}))
+|{\cal M}_{{\cal B}{\cal B'} \to \chi \chi}|^2 f^{\rm eq}_{\cal B}(\vec{k}) f^{\rm eq}_{{\cal B}'}(\vec{k'}) \nonumber\\
&\quad\times(1 \pm f_\chi(\vec{p})) (1 \pm f_\chi(\vec{p'}))
\Bigr),
\end{align}
\begin{align}
 C_{\rm el.}
&=
\frac{1}{2g_\chi}
\sum
\int \frac{d^3 p'}{(2\pi)^3 2 E_{p'}}
\int \frac{d^3 k}{(2\pi)^3 2 E_k}
\int \frac{d^3 k'}{(2\pi)^3 2 E_{k'}}
\nonumber\\
&\quad\times (2\pi)^4 \delta^4(p + p' - k - k') \nonumber\\
&\quad\times \Bigl(
-|{\cal M}_{\chi {\cal B} \to \chi {\cal B}}|^2 
f_\chi(\vec{p}) f^{\rm eq}_{\cal B}(\vec{k}) (1 \pm f_\chi(\vec{p'}))\nonumber\\
&\quad\times(1 \pm f^{\rm eq}_{\cal B}(\vec{k'})) 
+|{\cal M}_{\chi {\cal B} \to \chi {\cal B}}|^2 
f_\chi(\vec{p'}) f^{\rm eq}_{\cal B}(\vec{k'}) \nonumber\\
&\quad\times(1 \pm f_\chi(\vec{p})) (1 \pm f^{\rm eq}_{\cal B}(\vec{k}))
\Bigr),
\end{align}
where ${\cal B}$ and ${\cal B}'$ stand for the thermal bath particles, 
$g_\chi$ is the number of internal degrees of freedom of DM,
and $f_{\cal B}^{\rm eq}$ is given by the Fermi-Dirac or Bose-Einstein distribution depending on the spin of ${\cal B}$.
The summation should be taken for all the internal degrees of freedom
for all the particles.
For the nonrelativistic DM, $C_{\rm el.}$ can be simplified as the Fokker-Planck operator~\cite{Binder:2016pnr,Bertschinger:2006nq, Bringmann:2006mu, Bringmann:2009vf}: 
\begin{align}
\label{Eq:FP}
C_\mathrm{el} 
&\simeq
\frac{E}{2} \gamma(T)
{\Bigg [}
T E\partial_p^2 \!+ \left(  \frac{2 T E}{p} \!+\! p \!+\! \frac{T p}{E}\right) \partial_p + 3
{\Bigg ]}f_{\chi}.
\end{align}
where $T$ is the temperature of the thermal bath and $\partial_p\equiv\partial/\partial p$. 
In the above equation, the momentum transfer rate $\gamma(T)$ is given by (see also 
Ref.~\cite{Gondolo:2012vh}) 
\begin{align}
\label{eq:gamma}
\gamma&= \frac{1}{3 g_{\chi} m_{\chi} T} \! \int \! \frac{\text{d}^3 k}{(2\pi)^3} f_{\cal B}^{\pm}(E_k)\left[1\!\mp\! f_{\cal B}^{\pm}(E_k)\right] \int\limits^0_{-4 k_\mathrm{cm}^2} \! \! \!  \text{d}t (-t) \frac{\text{d}\sigma}{\text{d}t} v\,,
\end{align}
where $f_{\cal B}^{\pm}(E_k)$ is Fermi-Dirac or Bose-Einstein distribution of bath particle, $v$ is the relative velocity and $t$ is the Mandelstam variable. 
The differential cross section can be expressed as $({\text{d}\sigma}/{\text{d}t}) v$ $ \equiv$ $|\mathcal{M}|^2_{\chi {\cal B}\leftrightarrow\chi {\cal B}}$ /$(64 \pi {k} E_k m_\chi^2)$, 
and $k_{\rm cm}^2$ is given by 
\begin{equation}
 k_{\rm cm}^2 =\frac{m_\chi^2 ( E_k^2 - m_{\cal B}^2)}{m_\chi^2 + m_{\cal B}^2 + 2 m_\chi E_k}.
\end{equation}
Here $E_k$ is the energy of  heat bath particle ${\cal B}$.
Note that $k_{\rm cm}^2 \neq  E_k^2 - m_{\cal B}^2 = |\vec{k}|^2$.


During chemical decoupling, the scattering processes may no longer be frequent enough to maintain kinetic equilibrium, 
meaning that the DM particles acquire a temperature $T_\chi$ distinct from that of the thermal plasma in their subsequent evolution.
A common definition of the DM temperature is 
\begin{align}
 T_\chi (T)
&=
\frac{g_\chi}{3 n_\chi}\int \frac{d^3 p}{(2\pi)^3} \frac{\vec{p}^2}{E} f_\chi (\vec{p})\equiv
\frac{s^{2/3}}{m_\chi} y,
\end{align}
which is also a function of the thermal bath temperature $T$. In this definition $n_\chi$ is the number density of the DM,
and $s$ is the entropy density.  Here $y$ is a dimensionless version in analogy to the DM yield $Y (=n_\chi/s)$.

To properly describe the evolution of the DM temperature and subsequently explore the impact of EKD on the chemical decoupling process, 
we treat the second moment of $f_\chi$ as a dynamical degree of freedom.
By integrating Eq.~\eqref{Eq:boltzmann} with $g_\chi \int \frac{d^3 p}{(2\pi)^3} \frac{1}{E}$
and
$g_\chi \int \frac{d^3 p}{(2\pi)^3} \frac{1}{E} \frac{\vec{p}^2}{E^2}$, one obtains the zeroth and second moments of the Boltzmann equation, respectively. 
This leads to a relatively simple coupled system of Boltzmann differential equations, 
\begin{align}
\frac{Y'}{Y} &= \frac{s Y}{x \tilde H}\left[
\frac{Y_{\rm eq}^2}{Y^2} \left\langle \sigma v_{\mathrm{rel}}\right\rangle_T- \left\langle \sigma v_{\mathrm{rel}}\right\rangle_{T_\chi}
\right]\,, \label{aYfinalfinal}\\
\frac{y'}{y} &=   \frac{1}{x\tilde H}\langle C_{\text{el}} \rangle_2
+\frac{sY}{x\tilde H}\left[
\left\langle \sigma v_{\mathrm{rel}}\right\rangle_{T_\chi}-\left\langle \sigma v_{\mathrm{rel}}\right\rangle_{2,T_\chi}
\right]
+\frac{sY}{x\tilde H} \nonumber\\ 
&\quad\times\frac{Y_{\rm eq}^2}{Y^2}\left[
\frac{y_{{\rm eq}}}{y}\left\langle \sigma v_{\mathrm{rel}}\right\rangle_{2,T}\!-\!\left\langle \sigma v_{\mathrm{rel}}\right\rangle_T
\right]
+2(1-w)\frac{H}{x\tilde H}\,,\label{eq:acBE}
\end{align}
where $x$ is defined as usual $x = m_\chi/T$, 
$Y_{\rm eq}(x)\equiv n_{\rm eq}(T)/s$, and 
\begin{equation}
\tilde H\equiv H/\left[1+ (1/3)d(\log g^s_{\rm eff})/d(\log T)\right], 
\end{equation}
with $g^s_{\rm eff}$ being the entropy degrees of freedom of the 
background plasma. 
$w(T_\chi)\equiv 1-{\langle p^4/E^3 \rangle_{T_\chi}}/({6T_\chi})$, with $\langle p^4/E^3\rangle
=\frac{g_\chi}{n_\chi^{\rm eq}(T_\chi)}\int \frac{d^3p}{(2\pi)^3} {\left( \vec{p} \cdot \vec{p} \right)^2}/{E^3} e^{- \frac{E}{T_\chi}}$. 

The above compact form of the differential equations contains the following thermally averaged cross sections, 
\begin{align}
\langle C_{\text{el}} \rangle_2 \equiv \frac{g_\chi}{3 n T_{\chi}} \int \frac{\text{d}^3 p}{(2\pi)^3} \frac{p^2}{E^2} C_{\text{el}}\;,
\end{align}
\begin{align}
\VEV{\sigma v_{\mathrm{rel}}}_{T_\chi}
&\equiv
\frac{g_\chi^2}{(n_\chi^{\rm eq})^2}
\int \frac{d^3 p}{(2\pi)^3} 
\int \frac{d^3 q}{(2\pi)^3} 
\left( \sigma v_{\mathrm{rel}} \right)_{\chi \chi \to {\cal B} {\cal B}'}\nonumber\\
&\quad\times
f_\chi^{\rm eq}(\vec{p}, T_\chi)
f_\chi^{\rm eq}(\vec{q}, T_\chi). 
\end{align}
The thermal average $\langle \sigma v_{\mathrm{rel}}\rangle_{2,T}$ is a variant of 
the commonly used thermal average $\langle \sigma v_{\mathrm{rel}}\rangle_T$, and is explicitly stated in 
Ref.~\cite{Binder:2017rgn} and introduced as 
\begin{align}
\VEV{\sigma v_{\mathrm{rel}}}_{2, T_\chi}
&=
\frac{g_\chi^2}{(n_\chi^{\rm eq})^2 T_\chi}
\int \frac{d^3 p}{(2\pi)^3} 
\int \frac{d^3 q}{(2\pi)^3} 
\frac{\vec{p} \cdot \vec{p} }{3E} 
\left( \sigma v_{\mathrm{rel}} \right)_{\chi \chi \to {\cal B} {\cal B}'}\nonumber\\
&\quad\times
f_\chi^{\rm eq}(\vec{p}, T_\chi)
f_\chi^{\rm eq}(\vec{q}, T_\chi). 
\end{align}
For 
$\VEV{\sigma v_{\mathrm{rel}}}_{T} $
and 
$\VEV{\sigma v_{\mathrm{rel}}}_{2, T} $,
we replace $T_\chi$ by $T$ in 
$\VEV{\sigma v_{\mathrm{rel}}}_{T_\chi} $
and 
$\VEV{\sigma v_{\mathrm{rel}}}_{2, T_\chi} $, respectively.
Moreover, $n_\chi^{eq}(T_\chi)$ is given by
\begin{equation}
 n_\chi^{\rm eq}(T_\chi) = g_\chi \int \frac{d^3 p}{(2\pi)^3} f_\chi^{\rm eq}(\vec{p}, T_\chi)
= g_\chi \int \frac{d^3 p}{(2\pi)^3} e^{- \frac{E_p}{T_\chi}}.
\end{equation}

\bibliography{refs}

\clearpage

\end{document}